\documentclass[twocolumn,preprintnumbers,amsmath,amssymb,aps]{revtex4}

\usepackage[dvips]{graphicx}
\usepackage{dcolumn}
\usepackage{bm}
\usepackage{comment}
\usepackage{textcomp}
\usepackage{mathrsfs}

\def\XXint#1#2#3{{\setbox0=\hbox{$#1{#2#3}{\int}$}
\vcenter{\hbox{$#2#3$}}\kern-.5\wd0}}

\begin{document}


\draft



\title{Optical delay control of large-spectral-bandwidth laser pulses}

\author{Emilio Ignesti$^{1}$, Stefano Cavalieri$^{2,3}$, Lorenzo Fini$^{2,3}$, Emiliano Sali$^{2,3,*}$,\\
Marco V. Tognetti$^{1}$, Roberto Eramo$^{2,3,4}$ and Roberto
Buffa$^{1}$}

\affiliation{$^{1}$Dipartimento di Fisica, Universit$\grave{a}$
di Siena, Via Roma 56, I-53100 Siena, Italy.\\
$^{2}$Dipartimento di Fisica, Universit$\grave{a}$ di Firenze, Via
G. Sansone 1, I-50019 Sesto Fiorentino, Firenze, Italy.\\
$^{3}$European Laboratory for Non-Linear Spectroscopy (LENS),
Universit$\grave{a}$ di Firenze,\\
Via N. Carrara 1, I-50019 Sesto Fiorentino, Firenze, Italy.\\
$^{4}$INFM-CRS-Soft Matter (CNR), c/o Universit$\grave{a}$ la
Sapienza, Piazzale A. Moro 2, I-00185, Roma, Italy.\\
$^{*}$Corresponding author: sali@fi.infn.it
 }

\begin{abstract}
In this letter we report the first experimental observation of
temporal delay control of large-spectral-bandwidth multimode laser
pulses by means of electromagnetically induced transparency (EIT).
We achieved controllable retardation with limited temporal
distortion of optical pulses with an input spectral bandwidth of
3.3 GHz. The experimental results compare favorably with
theoretical predictions.
\end{abstract}


\maketitle 

\noindent In recent years the number of experiments aiming at the
reduction of the group velocity of laser pulses propagating
through material media has been growing steadily. The interest in
this research field concerns both fundamental issues (basic
physical laws of laser-matter interaction) and applications to
all-optical technologies. Many schemes have been proposed and
experimentally realized, based on effects such as
electromagnetically induced transparency (EIT)
\cite{Fleischhauer_Rev_77_633_2005}, coherent population
oscillations \cite{Bigelow_PRL_90_113903_2003,
Palinginis_Opticsexpress_13_9910_2005}, stimulated Brillouin and
Raman scattering \cite{Song_Optlett_30_1782_2005,
Stenner_Opticsexpress_13_9995_2005, Okawachi_PRL_94_153902,
Sharping_Opticsexpress_13_6092_2005,
Dahan_Opticsexpress_13_6234_2005}, spectral hole burning
\cite{Camacho_PRA_74_033801_2006} and double absorbing resonances
\cite{Camacho_PRA_73_063812_2006, Camacho_PRL_98_153601_2007}. The
possibility to coherently control not only the propagation
velocity of laser pulses, but also their temporal shape
\cite{Buffa_PRA_69_033815_2004}, probably makes systems based on
EIT the most promising for the design of all-optical delay lines
and buffers with very fast temporal switching. So far,
experimental observation of controllable temporal delay by means
of EIT has been reported only for narrowband
Fourier-transform-limited optical pulses, where the spectral
bandwidth $\delta \omega$ is inversely proportional to the pulse
temporal duration $\delta t$. However, such condition is not
strictly necessary in digital optical communications. Therefore,
in order to study the limitations of techniques based on EIT, it
appears of evident interest to investigate also the possibility of
temporal delay control of multimode non Fourier-transform-limited
laser pulses.

In this letter we report what, to the best of our knowledge, is
the first experimental observation of delay control by EIT of
large-spectral-bandwidth multimode dye laser pulses. We show how
EIT holds the potential to achieve controllable retardation of
laser pulses with a spectral bandwidth as large as 3.3 GHz.
\begin{figure}[hbt!]
  \includegraphics[width=8.4cm]{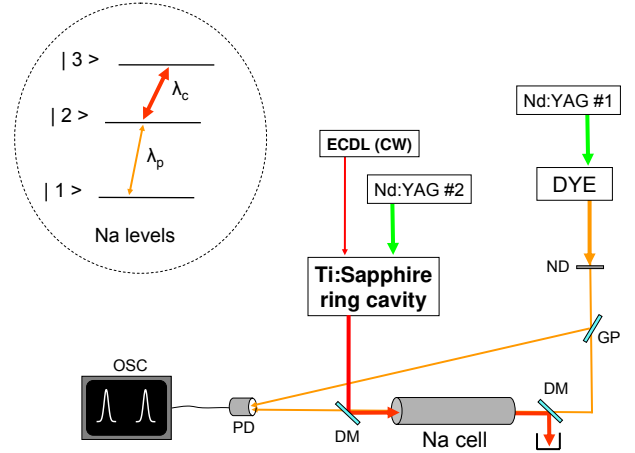}
  \caption{Experimental setup. ECDL: extended-cavity diode laser.
  ND: neutral-density filter. DM: dichroic mirror. GP: glass plate.
  PD: photo-diode. OSC: Digital oscilloscope. Top-left inset: scheme
  of sodium levels involved.}\label{Fig:setup}
\end{figure}

\begin{figure}
  \includegraphics[width=8.4cm]{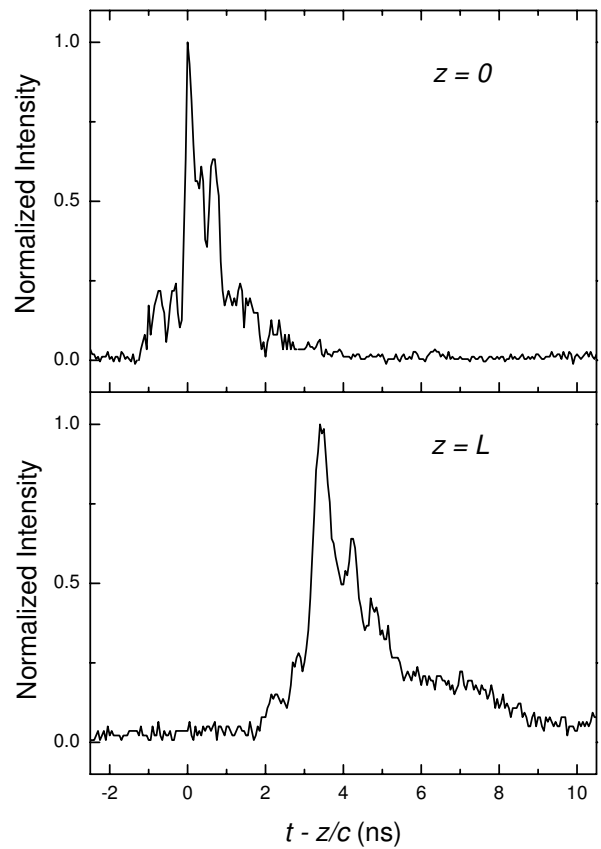}
  \caption{The upper and lower graphs show, respectively,
  the probe laser pulse at the input and at the output of the cell.
  Relative delay, transmission and likeness for this pulse are,
  respectively, $\Delta t = 3.4$ ns, $T=0.57$ and $\mathscr{L} = 0.90$).
  }\label{Figpulses}
\end{figure}

Figure \ref{Fig:setup} shows our experimental apparatus and, in
the top-left inset, the atomic levels and the transitions
involved. The three-level ladder scheme in sodium involves the
atomic states $|1\rangle=|2p^63s\,\,\,\,J=1/2\rangle$,
$|2\rangle=|2p^63p\,\,\,\,J=1/2\rangle$ and
$|3\rangle=|2p^63d\,\,\,\,J=3/2\rangle$. The probe field at
$\lambda_p \simeq 589.76$ nm, resonant with the transition
$|1\rangle$ - $|2\rangle$, is provided by a commercial
frequency-tunable multimode dye laser (Quantel TDL50) pumped by a
frequency-doubled Q-switched Nd:YAG laser at a repetition rate of
10 Hz. The dye laser pulses have a measured spectral bandwidth
$\delta \omega / 2\pi$ = 3.3 GHz and a multi-peak temporal
structure of several nanoseconds of duration. The temporal
evolution of a typical laser pulse is reported in Fig.
\ref{Figpulses}(a). The coupling field at $\lambda_c \simeq
818.55$ nm, resonant with the transition $|2\rangle$ -
$|3\rangle$, is provided by a home-made laser, since no
commercially available devices exist with the required
characteristics, i.e. frequency tunability,
single-longitudinal-mode operation (for temporal smoothness) and
adjustable pulse duration. The choice was that of a
titanium-sapphire (Ti:S) ring cavity pumped by another
frequency-doubled Q-switched Nd:YAG laser and injection-seeded by
a Toptica DL100 single-mode CW extended-cavity diode laser (ECDL)
\cite{raymond}. The single-mode emission of the ECDL can be tuned
in wavelength within the emission bandwidth of the diode laser
(typically a few nm). Both the wavelength of the CW injection-seed
from the ECDL and the wavelength of the pulsed emission from the
Ti:S cavity are monitored using a high-diffraction-order
spectrometer (Coherent Wavemaster) with an accuracy of 5 pm. The
sodium sample is contained in a cylindrical cell, heated up to
several hundreds degree Celsius for a length $L = 1$ m. All
measurements have been carried out at a temperature of 250$^\circ$
Celsius, corresponding to an independently measured value of the
density-length product of $2.0 \times 10^{15}$ cm$^{-2}$. The two
laser beams are linearly polarized along the same direction and
they overlap, both temporally and spatially, inside the cell. A
counter-propagating configuration was arranged in order for the
effect of Doppler broadening to be reduced. The temporal
synchronization of the laser pulses was obtained by mutually
adjusting the triggers of the two Nd:YAG pump lasers.
The length of the ring cavity of the control laser was chosen in
such a way to obtain a pulse temporal duration much longer than
that of the probe pulse. For a cavity length of 145 cm a temporal
duration full-width at half-maximum (FWHM) for the control pulse
of 62 ns was obtained. For such a value, the control field can be
considered sufficiently constant in time upon propagation of the
laser pulses along the cell. The probe beam has a slightly
elliptical profile with a 2-mm mean diameter, while the
overlapping control beam has a homogeneous circular profile with a
4-mm diameter.

\begin{figure}
  \includegraphics[width=8.4cm]{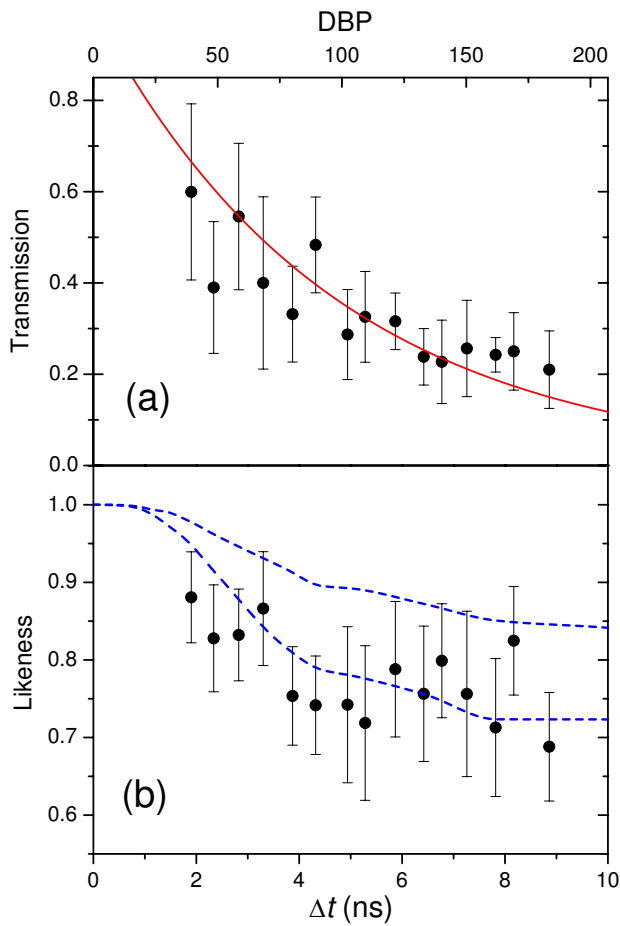}
  \caption{Transmission (a) and likeness (b) as a function
  of the pulse delay. The full circles with error bars are the experimental
  data. In (a) the continuous curve is a fit of the experimental data using
  the theoretical expression \ref{Eq:transpvsdelay}. In (b) the dashed lines
  represent the results of a theoretical calculation (see text for details).
  }\label{Figresults}
\end{figure}

All the measurements have been done with a probe pulse energy $E_p
\simeq 30$ nJ, corresponding to a peak intensity inside the cell
of approximately $I_p \sim 100$ W/cm$^2$, while, in order to
obtain various delays, we have varied the energy of the control
pulse from $E_c\simeq 0.3$ mJ to $E_c\simeq 7$ mJ, corresponding
to peak intensities inside the cell varying from $I_c\simeq 30$
kW/cm$^2$ to $I_c\simeq 700$ kW/cm$^2$. The detection system is
constituted by a fast photodiode (Antel Optronix PIN) and a
digital sampling oscilloscope (Tektronix TDS7704B, 7 GHz, 20
GSa/s). The overall temporal response of the system has been
measured to be Gaussian-shaped with a FWHM of 116 ps, when tested
with a sup-picosecond laser pulse. The value is sufficient to
resolve the temporal features of the probe pulse. As shown by
Fig.\ref{Fig:setup}, a fraction of the probe laser beam is
reflected from a glass plate before entering the cell, in order to
be used as a reference pulse to measure the delay. The reference
pulse is sent to the detection system together with the main part
of the probe pulse that exits the cell. The lengths of the two
optical paths are adjusted in order to be able to detect the two
pulses with the same detector without temporal overlapping. With
this configuration, several measurements of temporal delay and
energy absorption of the probe pulse were realized changing the
control laser intensity.

Figure \ref{Figpulses} shows an example of a comparison between a
pulse as it enters (top graph) and exits (bottom graph) the cell.
It can be seen that the main temporal features of the pulse are
preserved after propagation through the cell. A moderate
broadening of these features can be observed, indicating that the
spectral wings of the probe field are slightly affected by the
propagation through the delay line. In Fig. \ref{Figresults}(a)
the measured transmission $T$ is reported as a function of the
measured relative delay $\Delta t$ due to the effect of EIT. We
report pulse delays up to $\Delta t = 8.9$ ns. The experimental
data are in good agreement with the theoretical curve
\cite{kasapi}:

\begin{equation}\label{Eq:transpvsdelay}
T=exp(-2\gamma_{13} \Delta t)
\end{equation}

\noindent where $\gamma_{13}$ is the decoherence rate of the
two-photon transition 1-3. A best fit provides for the dephasing
rate the value $\gamma_{13} = 1.1 \times 10^8$ rad/s, which is
compatible with the combined contributions of radiative decay and
collisional effects due to residual gases in the cell.

An important parameter characterizing optical delay lines is the
\emph{delay-bandwidth product DBP}, given by the product of the
spectral bandwidth $\delta \omega$ of the optical field carrying
the information, times the temporal delay accumulated upon
propagation in the line: $DBP= \Delta t \times \delta \omega$. In
the top-horizontal axis of Fig. \ref{Figresults} we report the
very high values of $DBP$ that we obtain in our experiment (up to
a maximum of 183). Besides the $DBP$, another important parameter
is the temporal shape distortion suffered by the optical pulse.
For single-mode smooth pulses, the FWHM temporal width is an
adequate parameter to describe their temporal shape and the
broadening factor a commonly used parameter to characterize the
temporal distortion suffered upon propagation in the line. On the
contrary, for laser pulses with the multi-peak temporal structure
shown in Fig. \ref{Figpulses}, a much more adequate parameter is
the \emph{likeness} $\mathscr{L}$, defined as:

\begin{equation}\label{Eq:likeness}
\mathscr{L}(I_1,I_2)=1-\frac{\int_{-\infty}^{+\infty}
[I_1(t)-I_2(t)]^2\,\,dt}{\int_{-\infty}^{+\infty}
I_1^2(t)\,\,dt+\int_{-\infty}^{+\infty} I_2^2(t)\,\,dt}
\end{equation}

\noindent where for our case $I_1(t) = I_p(z=0,t)$ and $I_2(t) =
I_p(z=L,t-L/c-\Delta t)/T$. With this definition, $\mathscr{L}$
varies between 0 and 1 and its value for two identical pulses is
$\mathscr{L} = 1$. This quantity is reported as a function of the
relative delay $\Delta t$ in Fig. \ref{Figresults}(b), where the
experimental data (full circles) are compared with the results of
a numerical simulation. The numerical results are obtained by
integrating the probe pulse propagation equation for constant
coupling field in the presence of Doppler broadening
\cite{optlett}. The spectrum of the input probe field is generated
as the superposition of a series of Gaussian modes with a 3.3-GHz
wide Gaussian envelope. The modes are separated from each other by
0.5 GHz and each of them has a width of 0.15 GHz with a different
random constant phase. With these choices the temporal shape of
the obtained pulses reproduce quite well the main features of the
real pulses generated by the multimode dye laser. We observe that
the theoretical values of $\mathscr{L}$ for a given $\Delta t$ are
`dispersed' over a broad range due to the different spectral phase
of each pulse at the cell input. In order to show these
variations, we chose to report two curves (dashed lines in Fig.
\ref{Figresults}(b)), representing the values
$\mathscr{L}_m+\sigma_{\mathscr{L}}$ and
$\mathscr{L}_m-\sigma_{\mathscr{L}}$, where $\mathscr{L}_m$ and
$\sigma_\mathscr{L}$ are the mean value and the standard deviation
resulting from the propagation of 500 pulses with different
randomly generated initial spectral phases. The experimental data
are in satisfactory agreement with the theoretical calculations,
particularly so if we consider that the theoretical curves in Fig.
\ref{Figresults}(b) do not use any free parameter.

In summary, we presented the first experimental demonstration of
controllable delay of large-spectral-bandwidth multimode laser
pulses using EIT. We managed to delay by several nanoseconds and
with limited temporal distortion pulses with an input spectral
bandwidth of 3.3 GHz. These results provide useful information
about the potential of systems based on EIT for the purpose of
designing all-optical delay lines and buffers with very fast
temporal switching.

All the experimental work reported in this study was carried out
in the laboratories of the Department of Physics of the University
of Florence.

\end{document}